\documentstyle[preprint,aps,eqsecnum,epsf]{revtex}
\tighten

\def\OMIT#1{{ }}
\def\eqn#1{Eq.~(\ref{#1})}

\begin{document}

{\tighten
\preprint{\vbox{\hbox{DUKE-95-95}
\hbox{}
\hbox{}
\hbox{}
\hbox{}}}

{\tighten
\preprint{\vbox{\hbox{DUKE-95-95}
}}

\title{Hyperon Decays in Chiral Perturbation Theory Revisited}
}
\author{Roxanne P. Springer}
\address{Duke University Department of Physics, Durham, NC 27708
\\ {\tt rps@phy.duke.edu}}
\maketitle

\begin{abstract}
The discrepancies found between S-wave and
P-wave fits for hyperon decays are reinvestigated
using the heavy baryon
chiral Lagrangian formalism.  The
agreement is found to improve through the inclusion
of previously omitted diagrams.  The S-waves are
unaffected by this, but the P-wave predictions
are modified.  A correlated fit to the chiral
parameters is performed and the results discussed.
\end{abstract}

\bigskip
\date{August 1995}
\vfill\eject

\section{Introduction}

Chiral perturbation theory is an effective theory which obeys the
symmetries of QCD and contains a number of parameters which
must be determined experimentally.
If the theory reflects nature, then the parameters should
be universal.  This can be tested through one-loop SU(3)
breaking calculations for the decays of the
octet and decuplet baryons.  Uncalculable
terms may yield corrections of up to 30 percent to these
predictions, but if the variance in the comparison to
experiment goes beyond this,
the validity of the chiral expansion is questioned for that
process, and the reliability of estimating unmeasured
processes is open.
The two-body weak $\Delta s=1$ decays of hyperons are a natural place
to investigate the validity of chiral perturbation theory.
The experimental observables have been well measured,
and calculations including leading logarithmic corrections
(which appear through one-loop SU(3) breaking diagrams)
have been performed\cite{Wise,EJ}.  A comparison of these
results, however,
showed that the parameters which fit the S-wave
decays was inadequate for describing the P-wave decays.
This caused concern about the legitimacy of the chiral
Lagrangian expansion, at least for these processes\cite{Wise,Georgi}.
In this paper, previously
omitted diagrams have been included in the calculation of
P-wave amplitudes for nonleptonic hyperon decay.  A correlated
fit to the
three weak parameters is performed
and compared to previous results.  The fit to the data improves
markedly when one of the strong parameters, which
is not well constrained at present, is
also allowed to vary.

\section{The Chiral Lagrangian for Nonleptonic Decays.}

Heavy Baryon Chiral Perturbation Theory (HBChPT), which is
used to make predictions for hadronic processes
at momentum transfers much less than one GeV,
is introduced and well described in Ref.\cite{mj}.
The weak interaction portion of the Lagrangian
needed for $\Delta s=1$ hyperon decays, which
transforms under ${\rm SU(3)_L}\otimes  {\rm SU(3)_R}$
as an $(8_L,1_R)$,
is outlined in Ref.\cite{Wise,EJ}.  The Lagrangian
$${\cal L} = {\cal L}_{strong} + {\cal L}_{weak}$$
contains the particles which are dynamic in the
energy regime relevant for hyperon decay.  This includes
the lowest mass octet and decuplet of
baryons, and the octet of pseudo-Goldstone bosons.
\begin{eqnarray}\label{strong}
{\cal L}_{strong} &=& i \ {\rm Tr}\ \bar B_v\ \left(v\cdot {\cal D}
\right)B_v
+ 2\ D\  {\rm Tr}\ \bar B_v\ S_v^\mu\ \{ A_\mu, B_v \}
+ 2\ F\ {\rm Tr}\  \bar B_v\ S_v^\mu\ [A_\mu, B_v]
\nonumber \\
&&-\ i\ \bar
T_v^{\mu}\ (v \cdot {\cal D}) \  T_{v \mu}
+ \Delta m\ \bar T_v^{\mu}\ T_{v \mu}
+ {\cal C}\ \left(\bar T_v^{\mu}\ A_{\mu}\ B_v + \bar B_v\ A_{\mu}\
T_v^{\mu}\right){\phantom {f^2 \over 4}}
\nonumber\\
&& +\ 2\ {\cal H}\  \bar T_v^{\mu}\ S_{v \nu}\ A^{\nu}\  T_{v \mu}
+ {f^2 \over 8}\ {\rm Tr}\ \partial_\mu \Sigma \partial^\mu
\Sigma^\dagger
+ \mu\ {\rm Tr} \left( m_q\Sigma + m_q^\dagger\Sigma^\dagger \right) \
+\ \cdots \ \ \ \  ,
\end{eqnarray}
where $f \sim 93$ MeV is the meson decay constant, the light quark
mass matrix $m_q={\rm diag}\{m_u,m_d,m_s\}$, and
${\cal D_\mu}= \partial_\mu+[V_\mu,  \; ]$
is the covariant chiral derivative.  The subscript $v$ on the baryon
fields makes explicit that, in HBChPT, velocity is a good
quantum number and labels the field for this portion of the Lagrangian.
The actual full Lagrangian is a sum over all such velocities on terms
like that above.
The $B_v$ are the octet of baryons, and the $T_v^\mu$ are the
decuplet of baryons (the $\mu$ index is the Lorentz superscript for
this Rarita-Swinger field).
The vector and axial vector chiral currents used are defined by
\begin{eqnarray}
V_\mu&=&{1 \over 2} (\xi\partial_\mu\xi^\dagger +
\xi^\dagger\partial_\mu\xi)
\nonumber \\
A_\mu&=&{i \over 2} (\xi\partial_\mu\xi^\dagger -
\xi^\dagger\partial_\mu\xi)
\ \ \ .
\end{eqnarray}
Higher dimension operators, which contain
more derivatives or insertions of the light quark mass matrix,
are not needed in \eqn{strong} to the order we are working.
The octet of pseudo-Goldstone bosons, $M$, appears through
\begin{eqnarray}
\Sigma  = \xi^2= {\rm exp}\left( {2 i M\over f} \right) \ \ \ .
\end{eqnarray}

The strong couplings constants $F, D, {\cal C}$, and ${\cal H}$
have
been obtained by comparing one-loop computations of axial matrix
elements between octet baryons to semileptonic baryon decay
measurements\cite{mj}.
The constants ${\cal C}$ and ${\cal H}$ are further
constrained through the one-loop computation
of the strong decays of decuplet baryons \cite{bss}.
This yields,
\begin{eqnarray}\label{strongFDCH}
D =  0.6\pm 0.1& \hskip 2cm & F = 0.4\pm 0.1\ \     \nonumber \\
1.1 < |{\cal C}| < 1.8& \hskip 2cm &-2.8 < {\cal H} < -1.6
\ \ \ \ ,
\end{eqnarray}
\noindent Note that the sign of ${\cal C}$ remains a convention.
The errors do not include theoretical errors.

Assuming octet
dominance (the $\Delta I={1 \over 2}$ rule), the $\Delta s = 1$
weak Lagrangian is

\begin{eqnarray}\label{weak}
{\cal L}_{weak} &=&
G_F\ m_\pi^2\ \sqrt{2} f_\pi\  h_D\ {\rm Tr}\ {\overline B}_v\
\lbrace \xi^\dagger h\xi \, , B_v \rbrace \;
+ \;
G_F\ m_\pi^2\ \sqrt{2} f_\pi\ h_F\ {\rm Tr}\ {\overline B}_v\
{[\xi^\dagger h\xi \, , B_v ]} \; \nonumber \\ &&
+ G_F\ m_\pi^2\ \sqrt{2} f_\pi\ h_C\ {\overline T}^\mu_v\
(\xi^\dagger h\xi)\ T_{v \mu}  \; + \;
 G_F\ m_\pi^2\ h_\pi\ { f_\pi^2 \over 4}\ {\rm Tr} \left(  h \, \partial_\mu
\Sigma
\partial^\mu
\Sigma^\dagger  \right)
\ + \ \cdots\   ,
\end{eqnarray}
where
\begin{eqnarray}
h = \left(\matrix{0&0&0\cr 0&0&1\cr 0&0&0}\right)  \ \ \ ,
\end{eqnarray}

\noindent picks out just the $\Delta s=1$ piece needed for hyperon decays.
The constants $f_\pi$, $h_D, h_F, h_\pi$ and $h_C$ are
then fit to reproduce experimental data.  Predictive
power is obtained because there are many more observables
than parameters.
The pion decay constant $f_\pi \sim 93$ MeV.
Factors of $ G_Fm_\pi^2 \sqrt{2} f_\pi $ are inserted in \eqn{weak}\
so that the constants $h_D$, $h_F$, and $h_C$ are
dimensionless. Nonleptonic kaon decays suggest that the weak meson
coupling
$h_\pi = 1.4$.

\section{Hyperon Decay Amplitudes}

In this section, the formulae for the S-wave and P-wave amplitudes
for $\Delta S=1$ nonleptonic hyperon decay are discussed.  The
S-wave amplitudes were calculated previously\cite{EJ}.  The
portion of the P-wave amplitudes coming from the
diagrams in Figure \ref{P1}
were also calculated in \cite{EJ}.
The pieces which are new and the subject of this work affect the
P-wave amplitudes and arise from
the diagrams in Figure \ref{P2}.

The total amplitude for a decay of an initial octet
baryon to a final octet baryon, $B_i \rightarrow B_f \pi$
is given by
\begin{eqnarray}\label{spamp}
{\cal A} =  i\ G_F\ m_\pi^2\  \sqrt{2}\ f_\pi\ \overline{u}_{B_f} \  \left[
{\cal A}^{(S)}  +
 2 k \cdot S_v  {\cal A}^{(P)}  \right] u_{B_i} \ \ \ ,
\end{eqnarray}
where $k$ is the outgoing momentum of the pion and $S_v$ is
the spin operator for the baryons.
The amplitudes ${\cal A}^{(S)}$ and ${\cal A}^{(P)}$ are the
S-wave and
P-wave amplitudes.  Of all physically possible decays within
the octet of baryons, only four are independent after isospin
symmetry has been imposed.  In keeping with Refs. \cite{Wise,EJ},
we will continue to choose those four to be
$\Sigma^+ \rightarrow n \pi^+$,
$\Sigma^- \rightarrow n \pi^-$, $\Lambda \rightarrow p \pi^-$, and
$\Xi^- \rightarrow \Lambda \pi^-$.
The results will be given using the following definitions of
${\cal A}^{(S)}$ or ${\cal A}^{(P)}$:
\begin{eqnarray}
{\cal A}^{S,P}_{if}=\alpha^{S,P}_{if} + \left(\beta^{S,P}_{if} -
\lambda^{S,P}_{if}\ \alpha^{S,P}_{if}\right){m_K^2  \over 16\pi^2 f_K^2}
\log\left({m_K^2 \over
\Lambda_\chi^2}\right) \ \ \ ,
\end{eqnarray}
where
\begin{eqnarray}
\beta^{(P)}_{if}=\beta^{(P1)}_{if}+\beta^{(P2)}_{if} \ \ .
\end{eqnarray}
The kaon decay parameter and mass are $f_K$ and $m_K$, respectively,
and the chiral symmetry breaking scale, $\Lambda_\chi \sim 1$ GeV.

The $\alpha^{(S)}_{if}$, $\beta^{(S)}_{if}$ (including both
octet and decuplet intermediate states, denoted
$\overline{\beta}^{(S)}_{if}$ in Ref.\cite{EJ}), and
$\lambda^{(S)}_{if}$ terms can be found in Ref.\cite{EJ}.
Despite apparent differences in amplitude definitions, these
can be taken straight across because of the units used.  The
values finally obtained will be different simply because the fit to
parameters will include the changes in the P-wave amplitudes.

Similarly, the $\alpha^{(P)}_{if}$ and $\lambda^{(P)}_{if}$
are unaffected by the inclusion of the graphs in Figure \ref{P2}.
The $\overline{\beta}^{(P)}_{if}$ in Ref.\cite{EJ}
will now be called $\beta^{(P1)}_{if}$ and the new graphs will
give $\beta^{(P2)}_{if}$.  The diagrams in Fig. \ref{P2} yield
\begin{eqnarray}
\beta^{(P2)}_{\Sigma^+ n} &=& {D \over 3}
           {h_D +3h_F \over m_\Lambda-m_N} \ \lambda_\Lambda +
            F \ {h_F -h_D \over m_\Sigma-m_N}  \ \lambda_\Sigma -
            {(F+D)(h_F -h_D) \over m_\Sigma-m_N} \ \lambda_N \nonumber \\
\beta^{(P2)}_{\Sigma^- n} &=& {D \over 3} \
              {h_D +3h_F \over m_\Lambda-m_N} \ \lambda_\Lambda -
            F \ {h_F -h_D \over m_\Sigma-m_N} \ \lambda_\Sigma \nonumber \\
\beta^{(P2)}_{\Lambda p} &=&
         {2 D \over \sqrt{6}} \ {h_F -h_D \over m_\Sigma-m_N}
             \ \lambda_\Sigma -
            {F+D \over \sqrt{6}} \ {3h_F +h_D \over m_\Lambda-m_N}
           \ \lambda_N \nonumber \\
\beta^{(P2)}_{\Xi \Lambda} &=&
         - \ {D-F \over \sqrt{6}} \
           {3h_F -h_D \over m_\Xi-m_\Lambda} \ \lambda_\Xi +
            {2 D \over \sqrt{6}} \ {h_F +h_D \over m_\Xi - m_\Sigma}
             \ \lambda_\Sigma  \ \ \ ,
\end{eqnarray}
with
\begin{eqnarray}
\lambda_N &=& {17 \over 6}D^2-5DF+{15 \over 2}F^2+
       {1 \over 2}{\cal C}^2  \nonumber \\
\lambda_\Lambda &=&{7 \over 3}D^2+9F^2+{\cal C}^2  \nonumber \\
\lambda_\Sigma &=& {13 \over 3}D^2+3F^2+{7 \over 3}{\cal C}^2  \nonumber \\
\lambda_\Xi &=& {17 \over 6}D^2+5FD+{15 \over 2}F^2+
                       {13 \over 6}{\cal C}^2  \ \ .
\end{eqnarray}

\section{Discussion}

The amplitudes obtained from including the diagrams in Fig. \ref{P2}
are shown in Tables 1 and 2.  The experimental measurements, including
errors, are shown in the first column.  The second column contains
the tree level SU(3) predictions, where the chiral parameters
used are the ones extracted from tree-level fits.
The third column shows the results of Ref. \cite{EJ}.

The fourth column contains
the chiral one-loop predictions using weak parameters obtained
by fitting only to the S-wave experimental values.  To most
closely match the analysis of Ref. \cite{EJ}, the strong interaction
couplings are chosen to be F=0.4, D=0.61, $|{\cal C}|=1.6$, and
${\cal H}=-1.9$.  Letting the weak parameters $h_D$, $h_F$, and $h_C$
vary,
a fit to the S-wave decays yields \cite{min}
\begin{eqnarray}
h_D = -0.32 \pm 0.01, \hskip 1cm h_F = 0.98 \pm 0.03,  \hskip 1cm  h_C = -1.37
\pm 0.27 \ \ .
\end{eqnarray}
The errors shown are only those which arise from the experimental
variances.    The parameter
$h_C$ is not well determined and large variations in its value
do not appreciably change the predicted amplitudes.
The S-wave predictions are essentially unchanged using the
parameters above, and
the loop corrected chiral predictions are in excellent agreement
with experiment, as demonstrated in Ref. \cite{EJ}.
The situation for the P-wave
predictions is improved for $\Sigma^+ \rightarrow n \pi^+$,
where the agreement is within the allowed 30 percent
variation for chiral predictions.  For the decays
$\Sigma \rightarrow n \pi^-$
and $\Lambda \rightarrow p \pi^-$, the additional graphs
bring the prediction back to tree level values, while the $\Xi \rightarrow
\Lambda \pi^-$ decay remains essentially unchanged.

The fifth column in Tables 1 and 2 contains the results from using
both S-wave and P-wave amplitudes to fit the weak chiral parameters.
The tree level $\Omega$ decays for which there are experimental
results are used as well.  Expressions for these are in Ref. \cite{EJ}.
The strong decays of the decuplet favor midpoint values for
$|{\cal C}|$ and
${\cal H}$ of 1.2 and --2.2, respectively \cite{bss}.
A fit to $h_D$, $h_F$, and
$h_C$ in this scenario yields
\begin{eqnarray}
h_D = -0.38 \pm 0.01, \hskip 1cm h_F = 0.92 \pm 0.01,
\hskip 1cm  h_C = 0.74 \pm 0.18 \ \ .
\end{eqnarray}
The S-waves are still within 30
percent, but the P-waves get worse.
The nonleptonic hyperon decays clearly favor
a larger value for $|{\cal C}|$ than do the strong
decuplet decays.  The dependence on ${\cal H}$ is not
as sensitive.

Using the eight independent
nonleptonic hyperon decays, along with the $\Omega$ decays,
and ${\cal H}=-2.2$, the parameters ${\cal C}$, $h_D$, $h_F$,
and $h_C$ are allowed to vary.  The best fit is
obtained when
\begin{eqnarray}
|{\cal C}| = 1.76\pm 0.01&\hskip 2cm& h_D = -0.42\pm 0.01, \nonumber \\
 h_F = 0.76\pm 0.01&\hskip 2cm& h_C=0.26\pm 0.10
\end{eqnarray}
The matrix of correlation coefficients for this fit, given in the order
($h_C$, $h_D$, $h_F$, ${\cal C}$) is
\begin{eqnarray}
\left( \begin{array}{rrrr}
             1.000&-0.914&-0.974& 0.149 \\
           -0.914& 1.000 &0.919&-0.024 \\
           -0.974& 0.919& 1.000&-0.009 \\
          0.149&-0.024&-0.009& 1.000 \\
\end{array} \right)
\end{eqnarray}
The S-wave and P-wave amplitude predictions using
these parameters are given in the final column of each Table.
The S-waves remain well described, and
all but the $\Lambda \rightarrow p \pi^-$ P-wave modes
do as well as the S-waves.  This later decay amplitude
becomes positive for parameter values still within ranges
allowed by other observables, but the agreement
remains poor.
Still, the additional diagrams have improved
the situation to the point where the chiral expansion
appears to be on more solid footing with respect to the P-wave decays.
As Jenkins points out in Ref. \cite{EJ}, the large
corrections which the loop diagrams give to the tree-level
results need not be taken as evidence that the chiral
expansion is ill-behaved if it is the leading order terms
which are anomalously small rather than the loop effects
which are unnaturally large.

\vskip 1.0cm
\begin{tabular}{|| c || c | c | c | c | c | c ||}
\hline\hline
\rule{0cm}{0.5cm} & \multicolumn{6}{c||}{\em S-waves}
\\*[0.1cm] \cline{2-7}
\rule{0cm}{0.7cm} {\em decay}
& exp &  tree & theory\cite{EJ} & theory (S) & theory ($\Delta$) & theory
\\*[0.1cm]   \hline
\rule{0cm}{0.5cm}$\hspace{0.2cm} \Sigma^+ \rightarrow n \pi^+$
\hspace{0.2cm}
&\hspace{0.2cm} 	0.06 $\pm$ 0.01	\hspace{0.2cm}
&\hspace{0.2cm} 	0.00	\hspace{0.2cm}
&\hspace{0.2cm} 	--0.09	\hspace{0.2cm}
&\hspace{0.2cm} 	--0.09	\hspace{0.2cm}
&\hspace{0.2cm} 	0.00	\hspace{0.2cm}
&\hspace{0.2cm} 	--0.13	\hspace{0.2cm}
\\*[0.1cm]  \hline
\rule{0cm}{0.5cm}$\hspace{0.2cm} \Sigma^- \rightarrow n \pi^-$
\hspace{0.2cm}
&\hspace{0.2cm} 	1.88$\pm$0.01	\hspace{0.2cm}
&\hspace{0.2cm} 	1.21	\hspace{0.2cm}
&\hspace{0.2cm} 	1.90	\hspace{0.2cm}
&\hspace{0.2cm} 	1.88	\hspace{0.2cm}
&\hspace{0.2cm} 	1.74	\hspace{0.2cm}
&\hspace{0.2cm} 	1.90	\hspace{0.2cm}
\\*[0.1cm]  \hline
\rule{0cm}{0.5cm}$\hspace{0.2cm} \Lambda \rightarrow p \pi^-$
\hspace{0.2cm}
&\hspace{0.2cm} 	1.42$\pm$0.01	\hspace{0.2cm}
&\hspace{0.2cm} 	0.91	\hspace{0.2cm}
&\hspace{0.2cm} 	1.44	\hspace{0.2cm}
&\hspace{0.2cm} 	1.42	\hspace{0.2cm}
&\hspace{0.2cm} 	1.44	\hspace{0.2cm}
&\hspace{0.2cm} 	1.28	\hspace{0.2cm}
\\*[0.1cm]  \hline
\rule{0cm}{0.5cm}$\hspace{0.2cm} \Xi^- \rightarrow \Lambda \pi^-$
\hspace{0.2cm}
&\hspace{0.2cm} 	--1.98$\pm$0.01	\hspace{0.2cm}
&\hspace{0.2cm} 	--1.19	\hspace{0.2cm}
&\hspace{0.2cm} 	--2.04	\hspace{0.2cm}
&\hspace{0.2cm} 	--1.98	\hspace{0.2cm}
&\hspace{0.2cm} 	--1.91	\hspace{0.2cm}
&\hspace{0.2cm} 	--2.02	\hspace{0.2cm}
 \\*[0.1cm]  \hline\hline
\end{tabular}
\vskip 0.5cm
\parbox{6in}{Table 1. The S-wave $\Delta s=1$
hyperon amplitudes.  The first column is the experimental
result and the next is the tree level prediction of chiral
perturbation theory \cite{Wise,EJ}.  The third column
contains the loop corrected results of Ref.\cite{EJ}. The
``theory (S)'' column gives the fit using S-wave predictions only,
with the Ref. \cite{EJ} values
$|{\cal C}|$ = 1.6 and ${\cal H}$ = --1.9.
The ``theory ($\Delta$)'' column fits both S-wave and P-wave
expressions, but uses $|{\cal C}|$=1.2 and ${\cal H}$ = --2.2
taken from the strong
decuplet decays.
The last column uses the parameters which were obtained from
a best fit including both S-waves and P-waves, ${\cal H}$ = --2.2,
and ${\cal C}$ fit, including the diagrams
of Fig. \ref{P2}.}
\vfill\eject

\vskip 1.0cm
\begin{tabular}{|| c || c | c | c | c | c | c ||}
\hline\hline
\rule{0cm}{0.5cm} & \multicolumn{6}{c||}{\em P-waves}
\\*[0.1cm] \cline{2-7}
\rule{0cm}{0.7cm} {\em decay}
& exp &  tree & theory\cite{EJ} & theory (S) & theory ($\Delta$) & theory
\\*[0.1cm]   \hline
\rule{0cm}{0.5cm}$\hspace{0.5cm} \Sigma^+ \rightarrow n \pi^+$
\hspace{0.2cm}
&\hspace{0.2cm} 	1.81$\pm$0.01	\hspace{0.2cm}
&\hspace{0.2cm} 	--0.06	\hspace{0.2cm}
&\hspace{0.2cm} 	0.82	\hspace{0.2cm}
&\hspace{0.2cm} 	1.54	\hspace{0.2cm}
&\hspace{0.2cm} 	1.10	\hspace{0.2cm}
&\hspace{0.2cm} 	1.83	\hspace{0.2cm}
\\*[0.1cm]  \hline
\rule{0cm}{0.5cm}$\hspace{0.2cm} \Sigma^- \rightarrow n \pi^-$
\hspace{0.2cm}
&\hspace{0.2cm} 	--0.06$\pm$0.01	\hspace{0.2cm}
&\hspace{0.2cm} 	0.13	\hspace{0.2cm}
&\hspace{0.2cm} 	0.34	\hspace{0.2cm}
&\hspace{0.2cm} 	0.16	\hspace{0.2cm}
&\hspace{0.2cm} 	0.34	\hspace{0.2cm}
&\hspace{0.2cm} 	--0.04	\hspace{0.2cm}
\\*[0.1cm]  \hline
\rule{0cm}{0.5cm}$\hspace{0.2cm} \Lambda \rightarrow p \pi^-$
\hspace{0.2cm}
&\hspace{0.2cm} 	0.52$\pm$0.02	\hspace{0.2cm}
&\hspace{0.2cm} 	--0.28	\hspace{0.2cm}
&\hspace{0.2cm} 	--0.52	\hspace{0.2cm}
&\hspace{0.2cm} 	--0.27	\hspace{0.2cm}
&\hspace{0.2cm} 	--0.51	\hspace{0.2cm}
&\hspace{0.2cm} 	--0.11	\hspace{0.2cm}
\\*[0.1cm]  \hline
\rule{0cm}{0.5cm}$\hspace{0.2cm} \Xi^- \rightarrow \Lambda \pi^-$
\hspace{0.2cm}
&\hspace{0.2cm} 	0.48$\pm$0.02	\hspace{0.2cm}
&\hspace{0.2cm} 	0.11	\hspace{0.2cm}
&\hspace{0.2cm} 	0.35	\hspace{0.2cm}
&\hspace{0.2cm} 	0.34	\hspace{0.2cm}
&\hspace{0.2cm} 	0.67	\hspace{0.2cm}
&\hspace{0.2cm} 	0.48	\hspace{0.2cm}
 \\*[0.1cm]  \hline\hline
\end{tabular}
\vskip 0.5cm
\parbox{6in}{Table 2. The P-wave $\Delta s=1$
hyperon amplitudes.  The first column is the experimental
result and the next is the tree level prediction of chiral
perturbation theory \cite{Wise,EJ}.  The third column
contains the loop corrected results of Ref.\cite{EJ}. The
column labelled ``theory (S)'' uses the parameters obtained
from fitting to the S-wave expressions only, with
$|{\cal C}|$ = 1.6 and ${\cal H}$ = --1.9, and includes
the P-wave diagrams of Fig. \ref{P2}. The ``theory ($\Delta$)''
column fits both S-wave and P-wave decays, but uses
$|{\cal C}|$=1.2 and ${\cal H}$ = --2.2 taken from strong
decuplet decays.
The last column is the result of parameters extracted from a
best fit of both S-wave and P-wave expressions, with ${\cal H}$ = --2.2,
and ${\cal C}$ fit, including the diagrams
of Fig. \ref{P2}.}
\vskip 1.0cm

\section{Acknowledgements}

I would like to thank the Institute for Nuclear Theory at the
University of
Washington, where much of this work was completed,
for their kind hospitality. I gratefully acknowledge
advice from Martin Savage
and Ted Allen. I thank Martin for many discussions and
his always interesting
observations and suggestions, and Ted for invaluable
assistance with computers, codes, and error analysis.
This work is supported in part by the US Dept. of Energy under
grant number DE-FG05-90ER40592.

\begin{figure}
\epsfxsize=10cm
\hfil\epsfbox{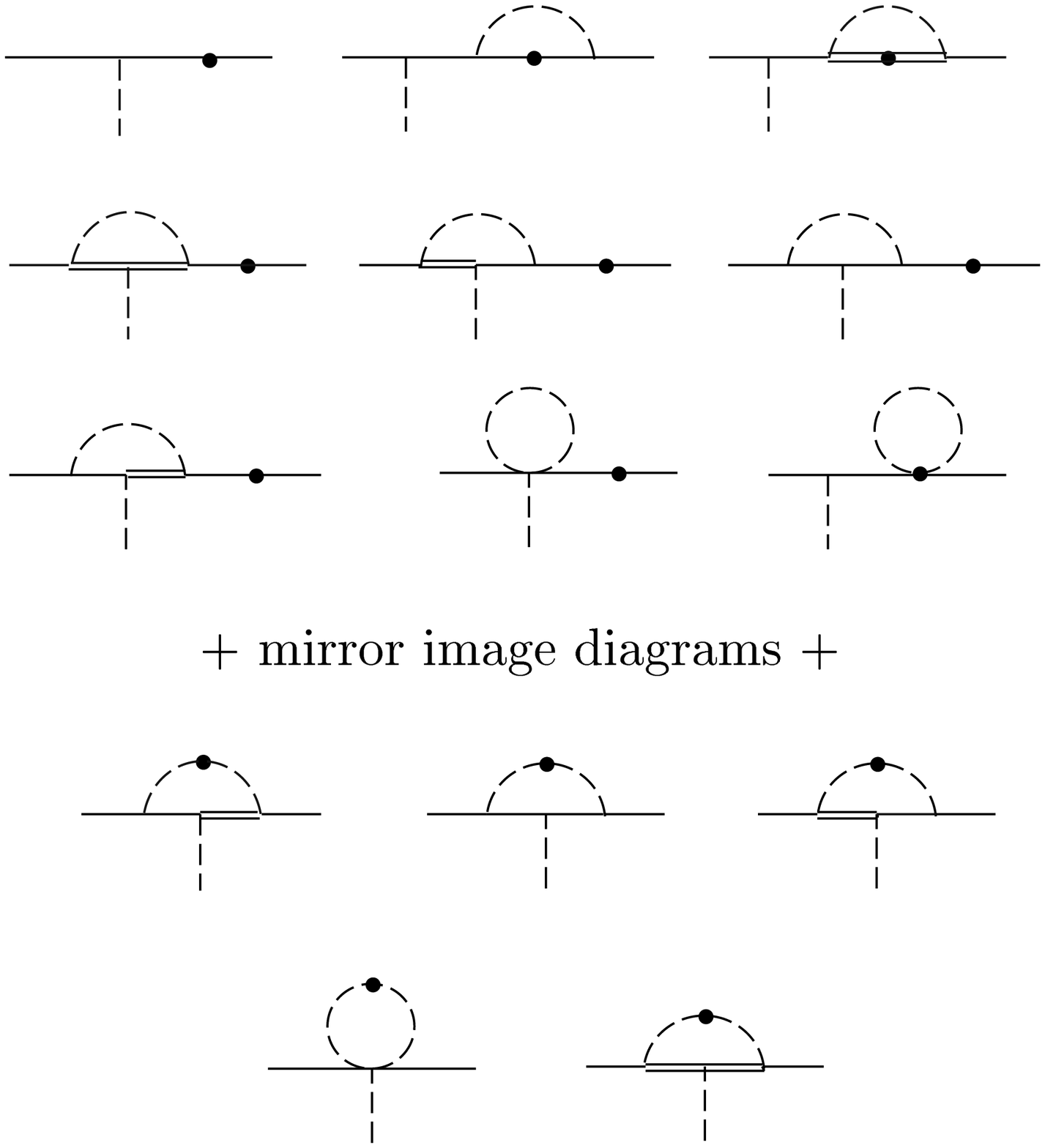}\hfill
\caption{Feynman diagrams for P-wave $\Delta s=1$ hyperon
decay calculated in Ref. [2].  The wavefunction renormalization
graphs are not shown.  The dashed lines
are mesons, and the solid lines are octet baryons.  The unmarked
vertex is a strong interaction and the black dots are weak vertices.}
\label{P1}
\end{figure}

\begin{figure}
\epsfxsize=10cm
\hfil\epsfbox{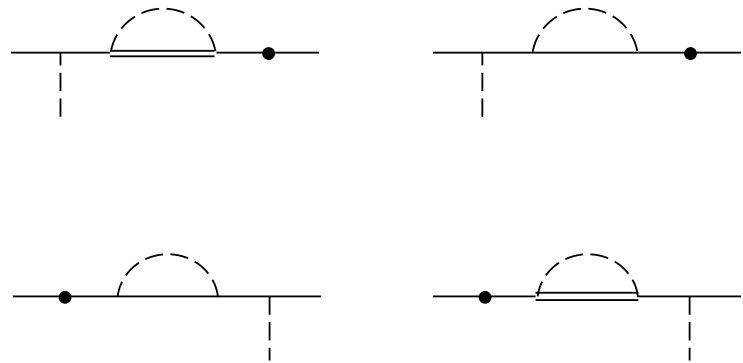}\hfill
\caption{Feynman diagrams for P-wave $\Delta s=1$ hyperon
decay included, in addition to the ones in Figure 1,
in the present calculation.
The dashed lines
are mesons, and the solid lines are octet baryons.  The unmarked
vertex is a strong interaction and the black dots are weak vertices.}
\label{P2}
\end{figure}

\end{document}